\documentclass[apjl,revtex4, apjfonts]{emulateapj}
\usepackage{amssymb,amsmath,mathrsfs,graphicx, url}
\usepackage[colorlinks=true, citecolor=blue, urlcolor=blue, linkcolor=blue]{hyperref}
\usepackage{verbatim}
\usepackage{subfigure}
\usepackage{xcolor}
\tabletypesize{\footnotesize}

\def\h2{$\rm H_2$}

\newcommand{\msun}{M$_{\odot}$}

\newcommand{\andii}{And~{\sc II}}
\newcommand{\andxvi}{And~{\sc XVI}}

\begin{document}

\title{Comparing M31 and Milky Way Satellites: The Extended Star Formation Histories of Andromeda {\sc II} and Andromeda {\sc XVI}\altaffilmark{*}}

\author{
Daniel R. Weisz\altaffilmark{1,2,3},
Evan D. Skillman\altaffilmark{4},
Sebastian L. Hidalgo\altaffilmark{5,6},
Matteo Monelli\altaffilmark{5,6},
Andrew E. Dolphin\altaffilmark{7},
Alan McConnachie\altaffilmark{8},
Edouard J. Bernard\altaffilmark{9},
Carme Gallart\altaffilmark{5,6},
Antonio Aparicio\altaffilmark{6,5},
Michael Boylan-Kolchin\altaffilmark{10}, 
Santi Cassisi\altaffilmark{11},
Andrew A. Cole\altaffilmark{12},
Henry C. Ferguson\altaffilmark{13},
Mike Irwin\altaffilmark{14},
Nicolas F. Martin\altaffilmark{15,16},
Lucio Mayer\altaffilmark{17,18},
Kristen B.~W. McQuinn\altaffilmark{4},
Julio F. Navarro\altaffilmark{19}, and
Peter B. Stetson\altaffilmark{8}
}

\altaffiltext{*}{Based on observations made with the NASA/ESA Hubble Space Telescope, 
obtained at the Space Telescope Science Institute, which is operated by the 
Association of Universities for Research in Astronomy, Inc., under NASA contract NAS 
5-26555. These observations are associated with program \#13028}

\altaffiltext{1}{Department of Astronomy, University of California at Santa Cruz,
1156 High Street, Santa Cruz, CA, 95064; drw@ucsc.edu}
\altaffiltext{2}{Astronomy Department, Box 351580, University of Washington, Seattle, WA, USA}
\altaffiltext{3}{Hubble Fellow}
\altaffiltext{4}{Minnesota Institute for Astrophysics, University of Minnesota, 
Minneapolis, MN, USA}
\altaffiltext{5}{Instituto de Astrof\'\i sica de Canarias. V\'\i a L\'actea s/n.
E38200 - La Laguna, Tenerife, Canary Islands, Spain}
\altaffiltext{6}{Department of Astrophysics, University of La Laguna. V\'\i a L\'actea s/n.
E38200 - La Laguna, Tenerife, Canary Islands, Spain}
\altaffiltext{7}{Raytheon; 1151 E. Hermans Rd., Tucson, AZ 85706, USA}
\altaffiltext{8}{Dominion Astrophysical Observatory, Herzberg Institute of 
Astrophysics, National Research Council, 5071 West Saanich Road, Victoria,
British Columbia V9E 2E7, Canada}
\altaffiltext{9}{Institute for Astronomy, University of Edinburgh, Royal
Observatory, Blackford Hill, Edinburgh EH9 3HJ, UK}
\altaffiltext{10}{Astronomy Department, University of Maryland, College Park, MD, USA}
\altaffiltext{11}{INAF-Osservatorio Astronomico di Collurania,
Teramo, Italy}
\altaffiltext{12}{School of Mathematics \& Physics, University of Tasmania,
Hobart, Tasmania, Australia}
\altaffiltext{13}{Space Telescope Science Institute, 3700 San Martin Drive, Baltimore, MD 21218, USA}
\altaffiltext{14}{Institute of Astronomy, University of Cambridge, Madingley Road, Cambridge CB3 0HA, UK}
\altaffiltext{15}{Observatoire astronomique de Strasbourg, UniversitŽ de Strasbourg, CNRS, UMR 7550, 11 rue de l'UniversitŽ, F-67000 Strasbourg, France}
\altaffiltext{16}{Max-Planck-Institut fŸr Astronomie, Kšnigstuhl 17, D-69117 Heidelberg, Germany}
\altaffiltext{17}{Institut f\"ur Theoretische Physik, University of Zurich,
Z\"urich, Switzerland}
\altaffiltext{18}{Department of Physics, Institut f\"ur Astronomie,
ETH Z\"urich, Z\"urich, Switzerland}
\altaffiltext{19} {Department of Physics and Astronomy, University of Victoria, BC V8P 5C2, Canada}

\begin{abstract}

We present the first comparison between the lifetime star formation histories (SFHs) of M31 and Milky Way (MW) satellites.  Using the Advanced Camera for Surveys aboard the Hubble Space Telescope, we obtained deep optical imaging of Andromeda {\sc II} (M$_{V} = -$12.0; log(M$_{\star}$/M$_{\odot}$) $\sim$ 6.7) and Andromeda {\sc XVI} (M$_{V} = -$7.5; log(M$_{\star}$/M$_{\odot}$) $\sim$ 4.9) yielding color-magnitude diagrams (CMDs) that extend at least 1 magnitude below the oldest main sequence turnoff, and are similar in quality to those available for the MW companions.  \andii\ and \andxvi\ show strikingly similar SFHs: both formed 50-70\% of their total stellar mass between 12.5 and 5 Gyr ago (z$\sim$5-0.5) and both were abruptly quenched  $\sim$ 5 Gyr ago (z$\sim$0.5).  The predominance of intermediate age populations in And {\sc XVI} makes it qualitatively different from faint companions of the MW and clearly not a pre-reionization fossil. Neither \andii\ nor \andxvi\ appears to have a clear analog among MW companions, and the degree of similarity in the SFHs of \andii\ and \andxvi\ is not seen among comparably faint-luminous pairs of MW satellites. These findings provide hints that satellite galaxy evolution may vary substantially among hosts of similar stellar mass. Although comparably deep observations of more M31 satellites are needed to further explore this hypothesis, our results underline the need for caution when interpreting satellite galaxies of an individual system in a broader cosmological context.

\end{abstract}


\section{Introduction}

Our current understanding of satellite and low mass galaxy evolution primarily comes from the Milky Way (MW) companions. Their close proximities (D $\lesssim$ 300kpc) enable a variety of detailed measurements including stellar abundances, radial and tangential velocities, stellar velocity dispersions, and deep resolved star color-magnitude diagrams (CMDs), which provide complementary constraints on their star formation and dynamical histories \citep[e.g.,][]{gallart2005, tolstoy2009, kirby2011, brown2012, sohn2013, weisz2014a}. Due to the availability of such detailed measurements, the MW companions are often used either explicitly or implicitly as the observable benchmarks for cosmological simulations of low mass galaxies and satellites systems \citep[e.g.,][]{munoz2009, busha2010, rocha2012, assmann2013, kazantzidis2013, starkenburg2013}. Considering their unique role as population templates, it is vitally important that we understand whether they are representative of satellites in the broader universe.

Testing the representative nature of the MW satellites requires comparison with other satellite populations. While deep imaging of the MW satellites is possible with ground-based telescopes \citep[e.g.,][]{sand2010, okamoto2012, delpino2013} most other systems are too distant for obtaining comparably detailed observations, making direct comparisons with the MW companions impossible. The exception is the M31 group. Its diversity of satellites and close proximity \citep[D$\sim$780kpc;][]{conn2012} allow for observations that approach the level of detail available in the MW companions, making it an excellent foil to the MW.

Currently, there is tentative evidence for systematic differences between the M31 and MW satellites. Several studies suggest that the M31 satellites follow a different size-mass relationship than the MW companions (e.g., \citealt{mcconnachie2006, kalirai2010}; although see \citealt{brasseur2011, tollerud2012} for an alternative interpretation) that may be related to systematically different dark matter profiles and/or a complex history of tidal interactions \citep[e.g.,][]{mayer2001, collins2014}. The sub-groups also different in their large scale structural properties. M31 hosts a rich set of streams, orphaned globular clusters, and a thin co-rotating plane of satellites \citep[e.g.,][]{mcconnachie2009, huxor2011, ibata2013, martin2013b}. In contrast, the MW sub-group appears to have fewer stream-like structures, orphan clusters, and a polar-oriented satellite configuration \citep[e.g.,][]{lyndenbell1976, belokurov2006b, pawlowski2012}. These structural differences are believed to trace the contrasting accretion histories of the two sub-groups \citep[e.g.,][]{shaya2013}.

Despite tantalizing differences in the present day satellite properties, little is known about the relative temporal evolution of the M31 and MW satellites. While the MW companions have ubiquitously deep CMDs and well-constrained SFHs \citep[$\lesssim$ 1 Gyr resolution at all ages; e.g.,][]{tolstoy2009, weisz2014a}, no comparable measurements have been made in the M31 group. Existing ground- and space-based imaging of the M31 satellites have only resulted in CMDs that include the horizontal branch, which are excellent for distance determinations, identifying new galaxies and clusters, and coarse stellar population characterization \citep[e.g.,][]{dacosta1996, dacosta2000,dacosta2002, mcconnachie2006, yang2012}, but are not suitably deep for well-constrained SFH measurements at all epochs \citep[e.g.,][]{gallart2005, weisz2014a}. As a result, we have little knowledge of major milestones in the M31 satellitesÕ histories such as the timing of the first epoch of star formation, the temporal patterns of stellar mass assembly, and the epochs of quenching, all crucial questions that have been answered for the MW companions through analysis of SFHs derived from deep CMDs \citep[][and references therein]{tolstoy2009, weisz2014a}.

In this paper, we undertake the first direct comparison of the lifetime SFHs of M31 and MW satellites. Using observations taken with the Advanced Camera for Surveys \citep[ACS;][]{ford1998} aboard the HST, we have measured the SFHs of two M31 companions, Andromeda {\sc II} and Andromeda {\sc XVI}, from CMDs that extend below the oldest main sequence turnoff (MSTO). The exceptional depth of these CMDs ensures that the resulting SFHs are directly comparable to the SFHs of MW companions, providing a first look at the temporal evolution of two satellite populations.

This paper is organized as follows.  We describe the observations, photometric reductions, and CMDs in \S \ref{sec:observations} and outline the SFH measurement method in \S \ref{sec:match}.  In \S \ref{sec:results} we present the SFHs of \andii\ and \andxvi, compare them with MW companion SFHs in \S \ref{sec:mwcomparison}.  We summarize our results in \S \ref{sec:conclusions}.  Throughout this paper, the conversion between age and redshift assumes the Planck cosmology as detailed in \citet{planck2013}.

\begin{deluxetable}{lll}
\tablecolumns{3}
\tablehead{
\colhead{Quantity} &
\colhead{And {\sc II}} &
\colhead{And {\sc XVI}} 
}

\startdata
(1) RA (J2000) &  01:16:27.0 & 00:59:29.8\\
(2) DEC (J2000) & $+$33:26:05.6 & $+$32:22:31.4\\
(3) (m-M)$_0$ & 24.07 $\pm$ 0.06  &  23.60 $\pm$ 0.2  \\
(4) M$_{V}$ & $-$12.0 $\pm$ 0.1 & $-$7.5 $\pm$ 0.3  \\
(5) A$_V$ & 0.17 & 0.18  \\
(6) r$_h$ (\arcmin) & 5.1 $\pm$ 0.1 &  0.93$_{-0.09}^{+0.16}$  \\
(7) Obs. Dates & Oct 4-6 2013 & Nov 20-22 2013   \\
(8) Orbits & 17 & 13 \\
(9) Exp. Time (F475W,F814W) (s) & 22472,17796  & 17194,13622   \\
(10) 50\% Comp. (F475W,F814W) & 28.8,27.9 & 28.8,27.8    \\
(11) Stars in CMD &80164  & 7695 
\enddata
\tablecomments{Basic observational properties of \andii\ and \andxvi. (1) - (3) are from \citet{mcconnachie2012}, (4) and (6) are from \citet{martin2014b}, (5) is from \citet{schlafly2011}.}
\label{tab:global}
\end{deluxetable}

\section{The Data}
\label{sec:observations}

\subsection{Observations and Photometry}

Our observational and data reduction strategy follow that of our previous program: Local Cosmology from Isolated Dwarfs (LCID).  Here, we briefly summarize that strategy and refer the reader to \citet{monelli2010b} for more details.

We obtained a single central field of HST/ACS imaging in both \andii\ and \andxvi\ between October 4 and 6 2013 and November 20 and 22 2013, respectively. In each galaxy, we also acquired an outer parallel field with the Wide Field Camera 3 \citep[WFC3;][]{kimble2008}.  In this paper, we focus on results from the central ACS fields, and will address the spatial dependences of the stellar populations in future papers.  Basic properties of both galaxies are listed in Table \ref{tab:global}.

We observed both galaxies in the F475W (Sloan g) and F814W (I).  The observations were taken over multiple orbits with a four step dither pattern in order to reject hot pixels and cosmic rays. The images were taken with a cadence that optimized the observations of short-period variable stars, which will also be presented in future papers.

We performed point spread function (PSF) photometry on the newly released charge transfer efficiency corrected images (i.e., \texttt{flc} images) for both galaxies using \texttt{DOLPHOT}, an updated version of \texttt{HSTPHOT} with an ACS specific module \citep{dolphin2000b}. Following the LCID strategy, we also performed photometry with \texttt{DAOPHOT}/\texttt{ALLFRAME} \citep{stetson1994}, and found no significant differences between the resulting CMDs, which is identical to the conclusions of the extensive photometric testing presented in \citet{monelli2010b} and  \citet{hidalgo2011}.  The remainder of this paper uses the \texttt{DOLPHOT} photometry.

From the raw photometric catalog, we rejected objects that did not meet particular requirements in signal-to-noise, PSF profile sharpness, and whose flux was significantly affected by neighboring objects.  Specifically, our accepted stars have SNR$_{\mathrm{F475W}}$ and SNR$_{\mathrm{F814W}}$ $>$ 5, (sharp$_{\mathrm{F475W}}$ + sharp$_{\mathrm{F814W}}$)$^2$ $<$ 0.1, and (crowd$_{\mathrm{F475W}}$ + crowd$_{\mathrm{F814W}}$) $<$ 1.0.  The precise definitions of these criteria can be found in \citet{dolphin2000b}.

To characterize the completeness and observational uncertainties, we inserted $\gtrsim 5\times 10^5$ artificial stars in the observed images and recovered their photometry in an identical manner to the real photometry. 

\subsection{Color-Magnitude Diagrams}
\label{sec:CMDs}

\begin{figure}
\begin{center}
\plotone{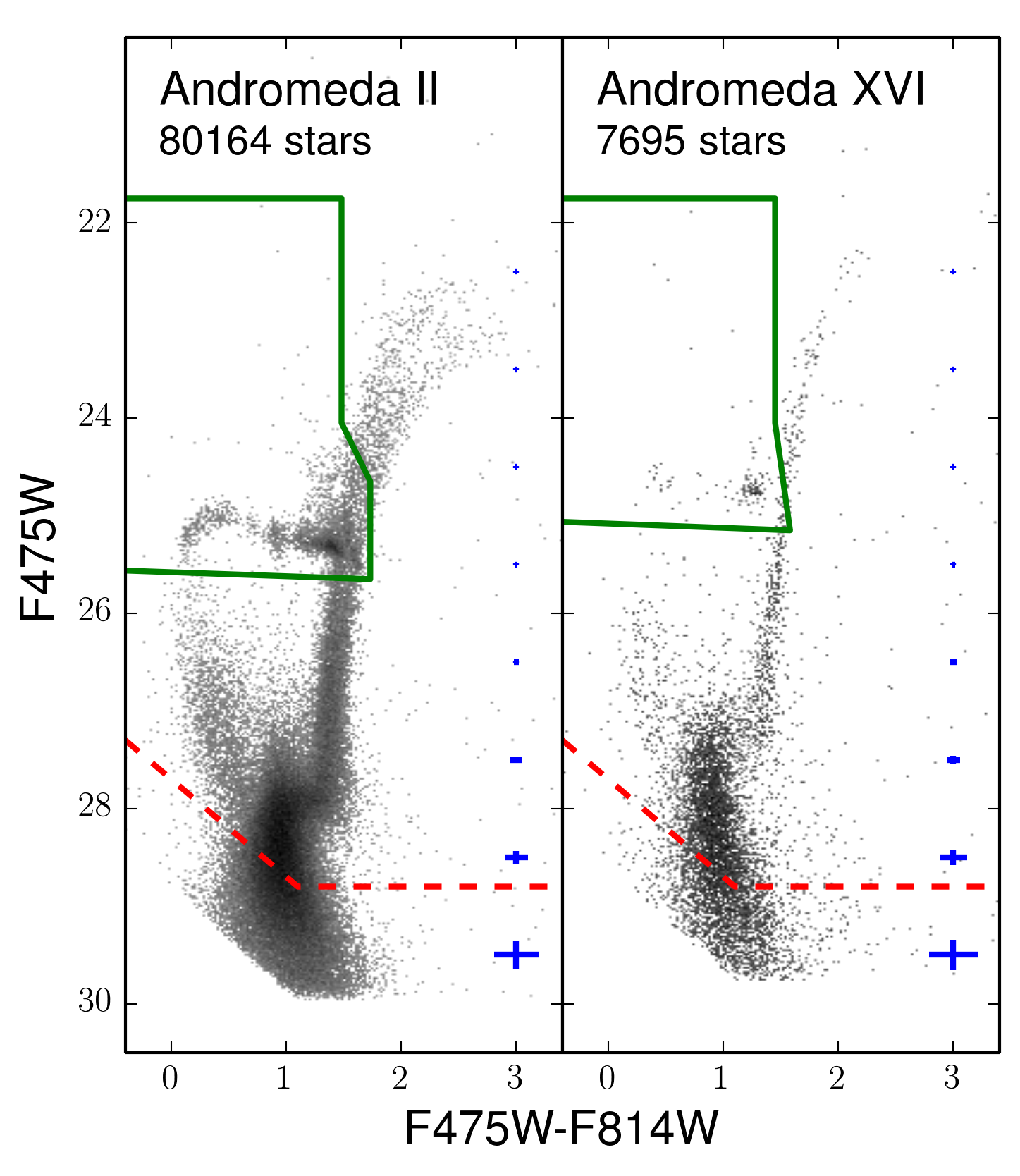}
\caption{The HST/ACS-based CMDs for \andii\ and \andxvi.  To enhance visibility of key CMD features over a large dynamic range of stellar densities we plotted these as Hess diagrams, i.e., finely binned CMDs.  The red-dashed lines reflect the 50\% completeness limits.  We excluded less certain phases of stellar evolution such as the horizontal branch and red clump from the CMDs fits, as indicated by the green polygons.}
\label{fig:cmds}
\end{center}
\end{figure}

In Figure \ref{fig:cmds}, we have plotted the CMDs of \andii\ and \andxvi.  In both systems, the photometry is 50\% complete to $\sim$ 1 mag below the oldest MSTO, providing for excellent leverage on the ancient SFHs of both systems.  The observations for both galaxies are several magnitudes deeper than any previous ground- or HST-based photometry of M31 satellites, making them the deepest observations ever obtained of satellite galaxies outside the virial radius of the MW. 

The two CMDs show several interesting features.  Most notably, \andii\ shows a split red giant branch (RGB), which indicates the presence of distinct age and/or metallicity populations.  It is the only known dwarf galaxy with a distinct double RGB.  Additionally, the CMD of And {\sc II} exhibits an extended red clump (RC), well-populated blue and red horizontal branches (HBs), and a sub-giant branch (SGB) that is broad in luminosity.

The CMD of \andxvi\ displays fewer distinct features and has lower stellar density, relative to \andii.  However, it too shows both a blue and red HB and an SGB that spans a brand range in luminosity, both of which may be indicative of an extended SFH.  We discuss the detailed SFHs of both galaxies in \S \ref{sec:results}.

\section{Measuring the Star Formation History}
\label{sec:match}

We have measured the SFHs of both galaxies using the CMD fitting package \texttt{MATCH} \citep{dolphin2002}.  Briefly, \texttt{MATCH} constructs a set of synthetic simple stellar populations (SSPs) based on user defined parameters such as a stellar initial mass function (IMF), age and metallicity bins, and searchable ranges in distance and extinction.  The synthetic SSPs are linearly combined and added to a model foreground population (from the empirical model in \citealt{dejong2010b}) to form a composite synthetic CMD, which is then convolved with observational biases from artificial star tests.  \texttt{MATCH} compares the model and observed CMDs using a Poisson likelihood statistic.  The SFH that corresponds to the best matched synthetic CMD is the most likely SFH of the observed population.   A full description of MATCH can be found in \citet{dolphin2002}.  The fitting of these CMDs uses the Padova stellar models \citep[][]{girardi2002, girardi2010} and follows the fitting methodology from \citet{weisz2014a}, with one exception: instead of fitting the full CMD, we excluded the red clump and HB from the fit in order to mitigate the contribution of these relatively less certain phases of stellar evolution to the SFH \citep{aparicio2009}. Throughout this paper, the plotted uncertainties reflect the 68\% confidence interval around the best fit SFH due to both random uncertainties (from a finite number of stars on the CMD) and systematic uncertainties (due to uncertain physics in the underlying stellar models).  We refer the reader to \citet{dolphin2012} for a full discussion of systematic uncertainties and \citet{dolphin2013} for a detailed description of random uncertainties in SFH measurements.

Following the LCID strategy, we tested the robustness of our SFHs by analyzing the data with a second CMD fitting package and a different set of stellar libraries.  In this case, we used \texttt{IAC-POP} \citep{aparicio2009} and the BaSTI stellar libraries \citep{pietrinferni2004} to measure the SFHs of both datasets.  Significant testing of the effects of excluding different parts of the CMD were also conducted \citep[cf.][]{monelli2010b}.  Overall, we found the solutions to be consistent within the plotted uncertainties and for the purpose of this paper, we will not discuss the details of these comparisons further and will use the \texttt{MATCH}-based SFHs.

\begin{figure}[ht!]
\begin{center}
\plotone{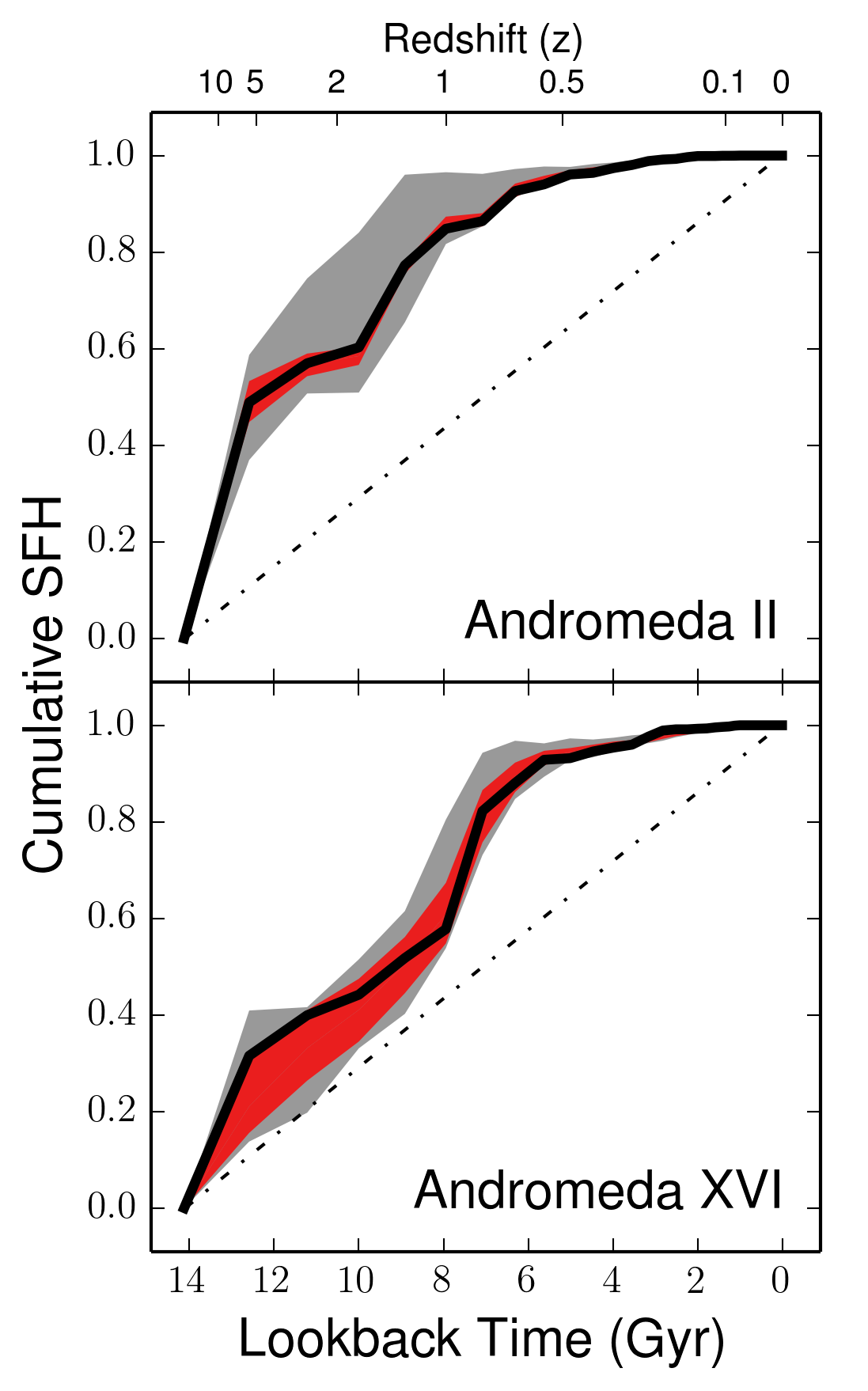}
\caption{The cumulative SFHs, i.e, the fraction of total stellar mass formed prior to a given epoch, of \andii\ and \andxvi.   The dot-dashed line reflects a constant lifetime SFR.  Random uncertainties are highlighted in color and the total uncertainties (random plus systematic) are shown in grey. The larger uncertainties for \andxvi\ are due to the smaller number of observed stars.}
\label{fig:andiisfh}
\end{center}
\end{figure}

\section{The Star Formation Histories of And {\sc II} and And {\sc XVI}}
\label{sec:results}

\begin{figure}
\begin{center}
\plotone{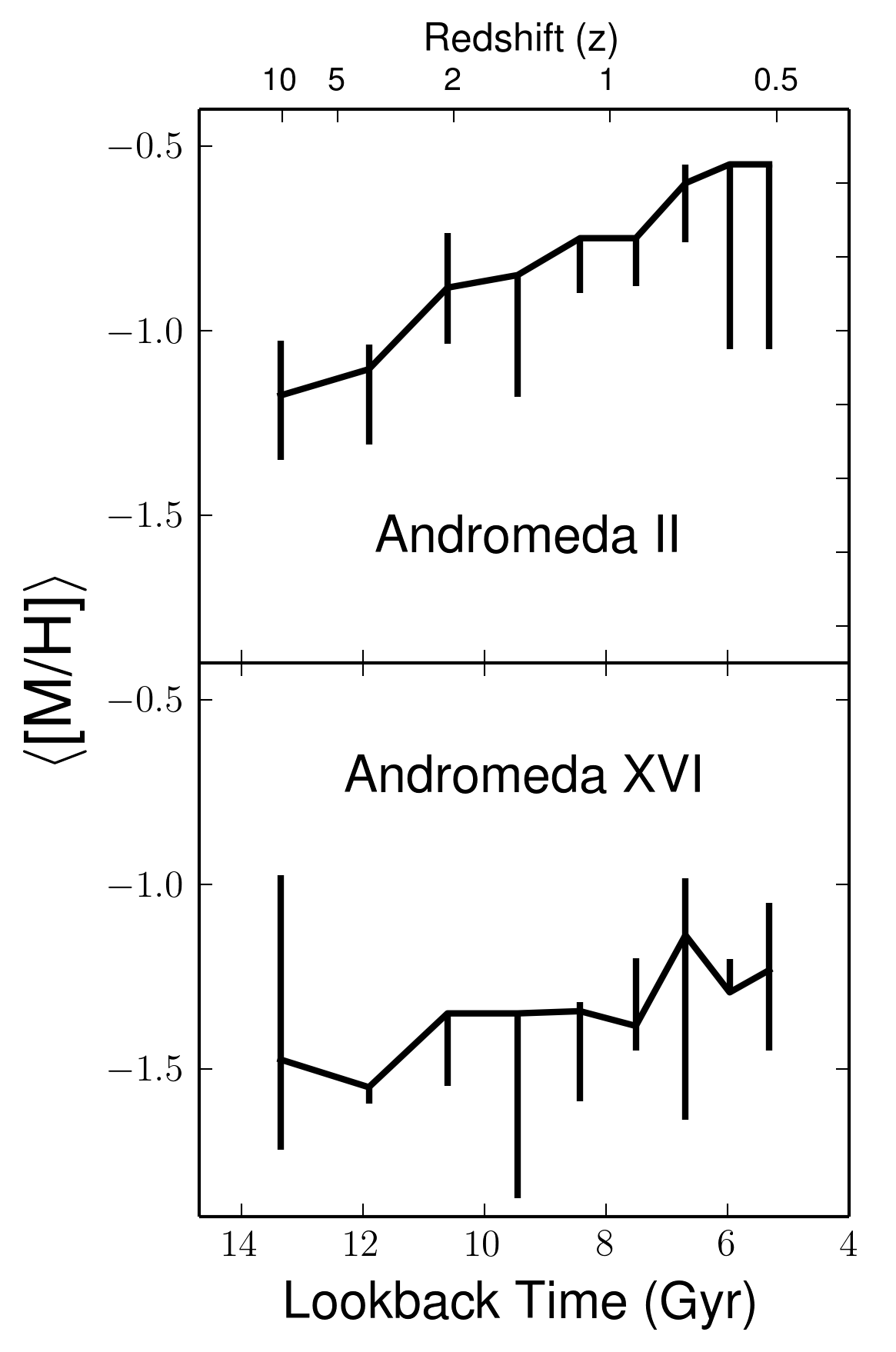}
\caption{The age-metallicity relationships of \andii\ and \andxvi over their intervals of active star formation.   The solid lines reflect the mean metallicity, and the error bars are the total uncertainties, i.e., random and systematic. .}
\label{fig:amr}
\end{center}
\end{figure}

In this paper, we focus on the cumulative SFHs, i.e., the fraction of stellar mass formed prior to a given epoch, which allow us to readily compare multiple SFHs on the same normalized scale, and are presented in Figure \ref{fig:andiisfh}.  For reference, the age-metallicity relationships (AMRs) for both galaxies are plotted in Figure \ref{fig:amr}. We will undertake a detailed interpretation of the AMRs and the absolute SFHs in future papers.

We first consider the SFH of \andii.  As shown in the top panel of Figure \ref{fig:andiisfh}, \andii\ formed $\sim$ 50\% of its total stellar mass prior to $\sim$ 12.5 Gyr ago (z$\sim$5) and $\sim$ 50\% of its stellar mass from 12.5-5 Gyr ago (z$\sim$5-0.5).  The initial burst was followed by a slower rate of mass growth from $\sim$ 12.5-10 Gyr ago (z$\sim$5-2), and an enhanced interval of star formation from $\sim$ 10-5 Gyr ago (z$\sim$2-0.5).  Our findings indicate that \andii\ had two distinct elevated periods of star formation, which will be discussed in detail in future papers.  Star formation in \andii\ was quenched at $\sim$ 5 Gyr ago (z$\sim$0.5).

Is it interesting to consider the SFH of \andii\ in light of its unusual properties.  From Subaru imaging, \citet{mcconnachie2007b} first noted that \andii\ has two distinct stellar populations: one centrally concentrated, metal-rich population and another extended, metal-poor population, similar to MW dwarfs such as Sculptor and Fornax \citep[e.g.,][]{tolstoy2004}.  However, \andii\ hosts a more spatially extended light profile than either Sculptor or Fornax \citep[e.g.,][]{mcconnachie2007b}.  Recently, \andii\ has been shown to both rotate about its minor axis \citep{ho2012} and host a kinematically cold stellar stream \citep{amorisco2014}, both of which are unique features among low mass galaxies.  Unfortunately, our CMD is entirely contained inside the stellar stream, making it challenging to directly tie our SFH to the merger scenario proposed by \citet{amorisco2014}.  In a future paper, we will leverage the wide-field ground based imaging along with our ACS and WFC3 observations to explore spatial variations in the populations of \andii, which may provide new insight into its unusual history.

The SFH of \andxvi, shown in the bottom panel of Figure \ref{fig:andiisfh} is similar to \andii, and hosts a mix of ancient and intermediate age populations.  \andxvi\ formed $\sim$ 30\% of its stellar mass prior to 12.5 Gyr ago (z$\sim$5) and 70\% of its mass between 12.5 and 5 Gyr ago (z$\sim$5-0.5).   Star formation in  \andxvi\ was also quenched at $\sim$5 Gyr ago (z$\sim$0.5).

The extended SFH of \andxvi\ is particularly intriguing in the context of cosmic reionization.  Several theoretical models predict that extremely low mass galaxies (M$_{\star}$ $\lesssim$ 10$^6$ \msun) should have had their star formation quenched $\sim$12.8-13.5 Gyr ago (z$\sim$ 6-14) due to heating of its gas by ultra-violet radiation from cosmic reionization \citep[so-called `fossils of reioniozation'; e.g.,][]{ricotti2005}.  Based on its low stellar mass (M$_{\star} \sim$ 10$^5$ \msun) and large distance from M31 (D$\sim$280 kpc), which reduces the role of environmental influence from M31, \andxvi\ is an ideal fossil candidate.  However, its large intermediate age population and continuous SFH strongly rule out reionization as a quenching mechanism.  

\begin{figure*}
\plotone{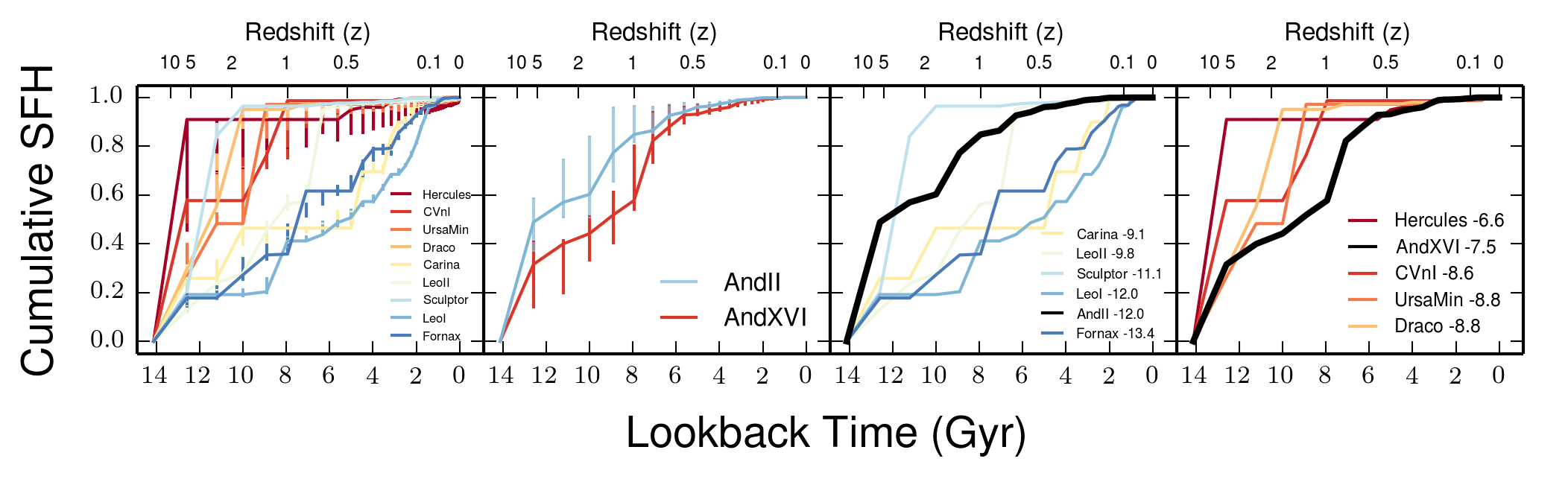}
\caption{A comparison between the SFHs of MW and M31 companions.  \textit{Left --}The SFHs of select MW companions from \citet{weisz2014a}.  Their properties are listed in Table \ref{tab:mwprops}. \textit{Left Middle --}  The SFHs of \andii\ and \andxvi, plotted identically. \textit{Right Middle --}  The SFH of \andxvi\ with comparable luminosity MW companions over plotted. \textit{Right  --}  The SFH of \andii\ with comparable luminosity MW companions over plotted.  Uncertainties have been omitted from the last two panels for clarity. }
\label{fig:m31_mw_comp}
\end{figure*}

Despite being nearly two orders of magnitude apart in stellar mass, \andii\ and \andxvi\ have extended SFHs that track one another remarkably well, as illustrated in Figure \ref{fig:m31_mw_comp}. Following the initial epoch of star formation,  both galaxies show declines in stellar mass growth beginning $\sim$ 12.5 Gyr ago, followed by brief increases in star formation activity, before finally being quenched at similar epochs of $\sim$ 5 Gyr ago (z$\sim$0.5).  While our small sample size cannot rule out chance coincidence, there are also speculative physical explanations for the similarity of their SFHs.  One possibility is that both galaxies may have similar halo masses, enabling them to retain gas for similar timescales.  The difference in stellar mass could potentially be attributed to large scatter in stellar mass at a fixed halo mass as discussed in \citet{boylankolchin2011} and \citet{garrisonkimmel2014}.  Another possibility is that both galaxies closely passed by M31 at similar times, but had different orbital trajectories that took them to their present day locations.  Interactions with massive hosts can transform gas-rich dwarf irregulars into gas-poor dwarf spheroidals in a few Gyr via `tidal stirring' \citep[][]{mayer2001}.

There is some evidence that other M31 satellites were quenched at intermediate epochs similar to And II and And XVI. Shallow CMDs of several other M31 satellites hint at extended SFHs that may have truncated several Gyr ago \citep[e.g., And {\sc I}, And {\sc III}, And {\sc XII};][]{dacosta1996, dacosta2000, dacosta2002, yang2012, weisz2014a}. However, the deepest existing CMDs only extend to the HB, prohibiting any statistically secure claims about quenching timescales; deeper photometry of more systems is needed to explore this scenario.

Intriguingly, well-constrained SFHs from deep CMDs of the M31 halo and outer disk also show strong declines $\sim$4-6 Gyr ago \citep[][]{brown2006, richardson2009, bernard2012}. Speculatively, the coincidental timing may be indicative of a global quenching event in the M31 sub-group perhaps due to a major merger in M31 at intermediate ages \citep[e.g.,][]{fardal2008}.  However, the detailed evolutionary relationship between a massive host and its satellites is not well-understood theoretically or empirically.

\section{A Comparison with the MW Satellites}
\label{sec:mwcomparison}

To facilitate a comparison between the MW and M31 satellite SFHs, we have plotted the cumulative SFHs of 10 MW companions (from \citealt{weisz2014a}) and our two M31 satellites in Figure \ref{fig:m31_mw_comp}.  We have selected the MW satellites that most resemble our two M31 galaxies in luminosity and current distance from their host galaxy.  Their properties are listed in Table \ref{tab:mwprops}.

Superficially, it appears that the MW companions and two M31 satellites may simply lie in a continuum of dSphs SFHs, with no regard to host galaxy properties. \andii\ and \andxvi\ both have some balance of ancient and intermediate age populations like most MW companions.

However, there are hints of important differences in this SFH comparison.  Most significant is the similarity in the SFHs of \andii\ and \andxvi, despite being two orders of magnitude apart in mass.   This is in stark contrast to similar pairings of the faint and luminous MW companions which do not show this same degree of similarity in SFHs, e.g., consider the SFHs of Hercules and Leo {\sc I}.  The uniformity in the SFHs of \andii\ and \andxvi\ and absence of similar pairs among the MW companions hints at the potential for unusual evolutionary behavior in the M31 group. This hypothesis can be further investigated with a larger sample of secure M31 satellite SFHs.

There are also some subtle differences when comparing individual galaxy SFHs.  \andii\ is similar in luminosity to MW companions such as Fornax and Leo {\sc I}.  However, its SFH is significantly different.  While Fornax and Leo {\sc I} both show constant SFHs until they were quenched $\sim$ 1 Gyr ago (z$\sim$0.1), \andii\ exhibits a qualitatively different SFH, before it was quenched 5 Gyr ago (z$\sim$0.5).   Sculptor is the next closest in luminosity, but it also exhibits a SFH different from that of \andii.  In terms of SFH, \andii\ bears some resemblance to Leo {\sc II}, which also was quenched around a similar time.  However, Leo {\sc II} had a dramatic burst of star formation $\sim$ 7 Gyr ago (z$\sim$0.7) before abruptly being quenched, while \andii\ formed stars steadily over intermediate ages.  As it stands, there appears to be no clear analog to \andii\ among the MW satellites.

\begin{deluxetable}{lccc}
\tablecolumns{4}
\tablecaption{Global Properties of Select M31 and MW Satellites}
\tablehead{
\colhead{Galaxy} &
\colhead{M$_{V}$} &
\colhead{M$_{\star}$} &
\colhead{Distance from Host}  \\
\colhead{} &
\colhead{} &
\colhead{(10$^6$ M$_{\odot}$)} &
\colhead{(kpc)} 
}
\startdata
\andii\ & $-$12.0 & 5.3 & 184 \\
\andxvi\ & $-$7.5 & 0.08 & 279 \\
Hercules & $-$6.6 & 0.04 & 126 \\
CVn {\sc I} & $-$8.6 & 0.23 &  218 \\
Ursa~Minor & $-$8.8 & 0.28 & 78 \\
Draco & $-$8.8 & 0.28 & 76 \\
Carina & $-$9.1 & 0.36 & 107 \\
Leo {\sc II} & $-$9.8 & 0.70 & 236 \\ 
Sculptor & $-$11.1 & 2.3 & 86 \\
Leo {\sc I} & $-$12.0 & 5.3 & 258 \\
Fornax & $-$13.4 & 19.0 & 149 
\enddata
\tablecomments{Luminosities and distances from nearest host for \andii, \andxvi, and select MW companions. All values from from \citet{mcconnachie2012}, except the luminosities (and stellar masses) of \andii\ and \andxvi, which are from \citet{martin2014b}.  Computation of the stellar masses assume \msun/L$_{\odot}$ $=$ 1 and a solar absolute V-band magnitude of 4.80.}
\label{tab:mwprops}
\end{deluxetable}

\andxvi\ also does not appear to have a counterpart in the MW subgroup.   \andxvi\ lies at the luminosity boundary between the so-called `ultra-faint' and `classical' dwarfs, but does not share a common SFH with members of either group.  The closest analogs are Canes~Venatici {\sc I} and Leo {\sc II}, which are $\sim$ 3 to 10 times more massive and located $\sim$ 50-80 kpc closer to the MW than \andxvi\ is to M31.  Leo {\sc II} also had an intermediate age burst before its quenching epoch, but it essentially experienced a constant SFH prior to the burst, which is different than \andxvi.

\andxvi\ is particularly unusual when compared to similarly low mass MW companions. The closest in mass is Hercules, which formed $>$ 90\% of its stellar mass $>$ 11 Gyr ago  \citep[e.g.,][]{brown2012}. Other, fainter MW satellites appear to have similarly old populations \citep[e.g.,][]{weisz2014a}. This comparison demonstrates that \andxvi\ is the lowest mass quenched galaxy known that hosts a predominantly intermediate age population. Leo T and Leo P are of similar stellar mass, but have cold gas and recent star-formation {weisz2012b, mcquinn2013}, making them qualitatively different than presently quenched satellites.

\section{Summary and Conclusions}
\label{sec:conclusions}

We have presented new HST/ACS-based CMDs of \andii\ and \andxvi\ that reach below the oldest MSTO, making them the deepest observations of satellite galaxies outside of the MW companions. From the deep CMDs, we derived their lifetime SFHs (with an age resolution $\lesssim$ 1 Gyr) that can be directly compared to SFHs of the MW satellites with minimal systematic effects.   \andii\ and \andxvi\ have similarly extended SFHs: both formed $\sim$ 50-70\% of their stellar mass prior from 12.5-5 Gyr ago (z$\sim$5-0.5), and were abruptly quenched $\sim$ 5 Gyr ago (z$\sim$0.5). This is particular striking as the galaxies are two orders of magnitude apart in stellar mass.  Among the MW companions, we find that neither \andii\ nor \andxvi\ have clear analogs, and that similar faint-luminous MW satellite pairings do not have such similar SFHs. Aside from chance coincidence, we discuss plausible physical scenarios to explain their similar SFHs including large scatter in the halo-stellar mass relationship and a global event in the M31 sub-group that may have affected the SFHs of multiple satellites.  The extended SFH of \andxvi\ strongly rules out quenching due to reionization, and makes it the lowest mass quenched galaxy (M$_{\star}$ $\sim$ 10$^5$ \msun) known to host a large intermediate age population.   While our findings hint at systematic differences between the M31 and MW satellites, similar quality observations of more M31 satellites are needed for further investigation.

\section*{Acknowledgements}

We would like to thank the anonymous referee for providing timely and insightful comments that improved the quality of the paper.  Support for this work was provided by NASA through grant number HST GO-13028 from the Space Telescope Science Institute, which is operated by AURA, Inc., under NASA contract NAS5-26555.  Support for DRW is provided by NASA through Hubble Fellowship grants HST-HF-51331.01 awarded by the Space Telescope Science Institute. This research made extensive use of NASA's Astrophysics Data System Bibliographic Services  and the NASA/IPAC Extragalactic Database (NED), which is operated by the Jet Propulsion Laboratory, California Institute of Technology, under contract with the National Aeronautics and Space Administration. In large part, analysis and plots presented in this paper utilized iPython and packages from Astropy, NumPy, SciPy, and Matplotlib \citep[][]{hunter2007, oliphant2007, perez2007, astropy2013}.

{\it Facility:} \facility{HST (ACS)}

\end{document}